# Quantized topological states and parity anomaly in intrinsic quantum anomalous Hall insulator MnBi2Te4

Zhongxun Guo[1,2,3,4,5†], Jingjing Gao[1,2,3,4,5†*], Zhiwei Huang[1,2,3,4,5], Di Yue[1,2,3,4,5], Zhaochen Liu[1,2], Mingyan Luo[1,2,3,4,5], Shuang Wu[1,2], Xinyu Chen[1,2], Guangyi Huang[1], Yujun Deng[1,2,3,4,5], Mengzhu Shi[6], Yin Xia[11], Zihan Xu[7], Chuanying Xi[8], Guangli Kuang[8], Changlin Zheng[1], Shiwei Wu[1,2], Hua Jiang[9], X. C. Xie[9,10], Wenzhong Bao[11], Yuping Sun[8,12,13], Xian Hui Chen[6*], Jing Wang[1,2*], Wei Ruan[1,2,3,4,5*], Yuanbo Zhang[1,2,3,4,5,14*]

[1]State Key Laboratory of Surface Physics and Department of Physics, Fudan University, Shanghai 200438, China

[2]Institute for Nanoelectronic Devices and Quantum Computing, Fudan University, Shanghai 200433, China

[3]Shanghai Branch, Hefei National Laboratory, Shanghai 201315, China

[4]Shanghai Research Center for Quantum Sciences, Shanghai 201315, China

[5]Zhangjiang Fudan International Innovation Center, Fudan University, Shanghai 201210, China

[6]Hefei National Laboratory for Physical Science at Microscale and Department of Physics, University of Science and Technology of China, and Key Laboratory of Strongly-coupled Quantum Matter Physics, Chinese Academy of Sciences, Hefei, Anhui 230026, China.

[7]SixCarbon Technology, Youmagang Industry Park, Shenzhen 518106, China.

[8]Anhui Province Key Laboratory of Low-Energy Quantum Materials and Devices, High Magnetic Field Laboratory, HFIPS, Chinese Academy of Sciences, Hefei 230031, China

[9]Interdisciplinary Center for Theoretical Physics and Information Sciences, Fudan University, Shanghai 200433, China

[10]International Center for Quantum Materials, School of Physics, Peking University, Beijing 100871, China;

[11]School of Microelectronics, Fudan University, Shanghai, 200433, China

[12]Key Laboratory of Materials Physics, Institute of Solid State Physics, Chinese Academy of Sciences, Hefei, China.

[13]Collaborative Innovation Center of Advanced Microstructures, Nanjing University, Nanjing, 210093, China.

[14]New Cornerstone Science Laboratory, Shenzhen 518054, China

[†] These authors contributed equally to this work.

[*] Correspondence should be addressed to Y.Z. (zhyb@fudan.edu.cn) and W.R. (weiruan@fudan.edu.cn), J.W. (wjingphys@fudan.edu.cn), X.H.C. (chenxh@ustc.edu.cn) J.G. (gaojj@fudan.edu.cn)

**When thinned down to just a few atomic layers, the layered magnetic topological insulator MnBi$_2$Te$_4$ offers an exceptional platform for exploring a wide range of topological phenomena[1–14]. In this work, we overcome longstanding challenges in synthesizing high-purity MnBi$_2$Te$_4$ crystals and report the observation of a myriad of quantized topological states in high-quality five-septuple-layer (5-SL) samples under magnetic fields up to 45 Tesla. We show that the nontrivial topology of 5-SL MnBi$_2$Te$_4$, in the presence of Landau quantization, is governed by a generalized topological index rooted in the parity anomaly of Dirac fermions in (2+1) dimensions[15–19]. The anomaly manifests as an anomalous Landau level, giving rise to gate-tunable helical edge transport. Our results establish high-quality MnBi$_2$Te$_4$ as a robust platform for exploring emergent topological states and for advancing novel quantum device applications.**

Over the past four decades, various quantized Hall effects have defined an ever-expanding frontier in condensed matter physics. Conventional quantum Hall effects rely on discrete, magnetic-field-induced Landau levels (LLs), which are essential for achieving exact quantization[20]. However, as first proposed by F. D. M. Haldane, a net external magnetic field is not always necessary—a system with a nontrivial topological electronic structure can defy the LL paradigm and exhibit quantized Hall conductance even at zero external magnetic field[15]. This insight has ultimately led to the experimental observation of the zero-field quantum anomalous Hall (QAH) effect in magnetically doped topological insulators, chromium- and vanadium-doped (Bi,Sb)$_2$Te$_3$[21–25], which open exciting opportunities for compact precision metrology[26–31], low-dissipation electronics[21,32–34], and the realization of chiral Majorana edge states for topological quantum computing[35–39].

Realizing the full potential of the magnetically doped topological insulators, however, remains a significant challenge. The short-ranged exchange interaction between the magnetic dopants requires a high doping level (typically around 15%) to establish long-range magnetic order, but these randomly distributed dopants also act as impurities that degrade the quantum transport in this class of materials. The recent advent of MnBi$_2$Te$_4$[1–6], a topological material with innate magnetic order, raises hope for overcoming this dilemma without the need for magnetic doping or additional band structure engineering[40–46]. Indeed, zero-field Hall conductance quantization has been achieved at a temperature of $T = 1.4$ K in mechanically exfoliated MnBi$_2$Te$_4$ flakes with 5 SLs[7], and later reproduced in 7-SL flake and epitaxially-grown MnBi$_2$Te$_4$ films[47–49]. Furthermore, theoretical calculations for ideal, pristine MnBi$_2$Te$_4$ anticipate a topological energy gap on the order of 88 meV—the highest across all existing QAH systems[34]. These developments highlight MnBi$_2$Te$_4$'s tremendous potential as a platform for exploring exotic topological quantum phenomena and potential applications in topological electronics, while also revealing that sample quality remains a major bottleneck limiting further progress with MnBi$_2$Te$_4$.

In this work, we significantly enhance the quality of MnBi$_2$Te$_4$ samples by synthesizing high-purity bulk crystals and optimizing sample fabrication. The improved sample quality is demonstrated by a substantial QAH gap up to 30 K, with its field-dependence allowing us to delineate the magnetic structure in the MnBi$_2$Te$_4$ flakes. More importantly, the high sample quality enables the electronic transport to reach the extreme quantum limit under magnetic fields up to 45 T, creating a unique opportunity to probe topological states arising from both Landau level (LL) quantization and intrinsic topological order. Remarkably, we find that the sequence of quantized states observed in our experiment defies conventional LL indexing. Instead, these topological states are characterized by a generalized topological index, a concept fundamentally linked to the parity anomaly of Dirac fermions in (2+1) dimensions[15–19].

Through model calculations, we identify an anomalous LL that reflects the parity anomaly under high magnetic fields. These calculations further predict that this anomalous LL supports a distinctive helical edge transport, which we confirm experimentally.

We begin by synthesizing high-purity single crystals of bulk $MnBi_2Te_4$ using a refined self-flux method. A key finding is that replacing the commonly used quenching process by slow cooling to room temperature effectively suppresses the formation of $Mn_{Bi}$ anti-site defects, yielding high-quality bulk $MnBi_2Te_4$ crystals (see Methods; the X-ray diffraction (XRD) and transmission electron microscopy (TEM) characterization of the crystals is presented in Supplementary Fig. 1). We then mechanically exfoliate the single crystal into atomically thin flakes and pattern metal contacts on top by evaporating Au/Cr (with typical thickness of 70 nm/2 nm) through stencil masks aligned with the flakes. Finally, we use a sharp tip to trim any excess $MnBi_2Te_4$, producing devices with uniform thicknesses and well-defined edges that are suitable for subsequent transport measurements. This study focuses on 5-SL specimens in two device configurations. The first, referred to as "Type A", involves 5-SL $MnBi_2Te_4$ flakes exfoliated using an $Al_2O_3$-assisted method as described in ref.[50]. Here we further streamline the exfoliation process by pressing the $Al_2O_3$-coated bulk crystal directly onto a sticky substrate covered with a thin adhesive layer from thermal release tape, eliminating heating steps that would otherwise degrade sample quality[51] (Supplementary Fig. 2). The second device type, "Type S", employs $MnBi_2Te_4$ 5-SL flakes directly exfoliated onto $SiO_2$/Si wafers. Given the high sensitivity of $MnBi_2Te_4$ flakes to oxygen and moisture, all fabrication processes are conducted in an argon-filled glove box with $O_2$ and $H_2O$ levels maintained below 0.1 parts per million to minimize sample degradation.

Fig. 1a presents a schematic of a Type A device, featuring an actual 5-SL $MnBi_2Te_4$ flake (Device A1) supported on a transparent substrate. The substrate consists of a 50-nm-thick indium tin oxide (ITO) layer and a 100-nm-thick $Al_2O_3$ layer, sequentially grown on a quartz (1000) wafer using atomic layer deposition (ALD). This fully transparent substrate enables accurate layer identification of $MnBi_2Te_4$ flakes via transmittance imaging[7], while the conductive ITO layer functions as a back gate for tuning the carrier density in the flakes. Fig. 1b displays the temperature-dependent longitudinal sample resistance, $R_{xx}(T)$, of four 5-SL Type A devices. As temperature $T$ approaches zero, all $R_{xx}(T)$ curves exhibit a sharp drop, a significant departure from the rapid increase reported in previous devices[7]. The sharp drop signifies the emergence of dissipationless edge transport, indicating a much-enhanced device quality.

Indeed, these devices demonstrate substantially improved QAH quantization. Fig. 1c displays the quantized Hall resistance $R_{yx}$ alongside the vanishing longitudinal resistance $R_{xx}$, measured in a typical 5-SL Type A device (Device A1) at $T = 1.5$ K as the back-gate voltage $V_g$ tunes the device near the charge neutrality point (CNP) of $V_{CNP} \approx -20$ V. The quantization of $R_{yx}$ achieves a precision better than 1 part in $10^4$, within our measurement uncertainty after correcting for the systematic errors. The quantization remains within 97% of the unit quantum resistance $h/e^2$ (where $h$ is the Planck constant and $e$ is the elementary charge) up to $T = 4.0$ K (Fig. 1d upper panel). An external magnetic field induces a sharp ferromagnetic (FM) transition, reversing the sign of $R_{yx}$. Notably, our improved crystal growth and device fabrication methods yield a $\sim 50\%$ success rate (6 out of 11 devices) for observing the zero-field QAH effect.

The improved device quality is further reflected in the elevated energy scale of the QAH

observed. Specifically, we extract the QAH energy gap $\Delta E$ from the slope of a line fit to the Arrhenius plot of thermally-activated $R_{xx}$, where $R_{xx} \propto \exp(-\Delta E / 2k_B T)$, and $k_B$ is the Boltzmann constant. This fit yields a zero-field QAH gap as high as $\Delta E = 29.0$ K (Fig. 1e, Device A1)—three times greater than previously reported for few-layer $MnBi_2Te_4$ and comparable to the largest gaps obtained across all QAH systems[7,21–25,40–43,45–48].

More importantly, measurements of the QAH gap under varying magnetic fields provide key insights into the nature of magnetic transitions in Type A devices. These transitions are evident in the magnetic field dependence of $R_{xx}$ measured at elevated temperatures, as marked by colored ticks in Fig. 1e. Arrhenius plots of $R_{xx}$ yield $\Delta E$ as a function of magnetic field $\mu_0 H$, with each magnetic state exhibiting distinct field-dependent behavior (Fig. 1e). At low magnetic fields ( $\mu_0 H \leq \mu_0 H_1 = 2.2$ T ), the interlayer antiferromagnetic (AFM) state dominates, characterized by a $\Delta E$ that decreases linearly with increasing field. The slope of this linear trend corresponds to an effective $g$-factor of $-2.1$, consistent with Zeeman narrowing of the QAH gap as the external magnetic field counteracts the exchange field experienced by electrons near the Fermi level[52]. At high fields, a canted antiferromagnetic (CAFM) state develops at $\mu_0 H_2 \approx 4.8$ T, eventually transitioning to a fully polarized FM state at $\mu_0 H_3 \approx 8.4$ T[53,54].

The magnetic state in the intermediate field range, $\mu_0 H_1 < \mu_0 H < \mu_0 H_2$, where $\Delta E$ exhibits a peak, is particularly intriguing. While signatures of this state were previously observed in $AlO_x$-capped few-layer samples[7,47,49], it was absent in most other measurements. We attribute this state to the collinear magnetized $M3'$ state predicted by both classical Monte Carlo simulations[55] and AFM linear-chain models[47] (Supplementary Fig. 4). This $M3'$ state features significantly enhanced magnetization at both surfaces (Fig. 1e inset), which explains the drastic increase in $\Delta E$. The similarities in both the field-dependent QAH gap behavior and the simulation results between our system and $AlO_x$-capped 7-SL MBT device[47] suggest that they share a common magnetic origin. Our simulations further reveal the nominal parameters of $MnBi_2Te_4$ place the system near a critical point in the magnetic phase diagram, where slight variations in defects and disorder can lead to substantial sample-to-sample differences observed across various experiments[7,47,49,54,56–58]. Indeed, Type A devices exhibit a similar set of magnetic transitions, albeit at magnetic fields distinct from those in Type S devices (Fig. 1f; see the schematic of Type S devices in Fig. 2c lower inset).

Furthermore, we find that defects and disorder strongly influence the QAH quantization in $MnBi_2Te_4$. Specifically, we observe a pronounced anti-correlation between $Mn_{Bi}$ (Mn residing on the Bi site) anti-site defect density and remnant zero-field $R_{yx}$ in Type A devices—higher defect densities correspond to lower remnant $R_{yx}$ (Supplementary Fig. 3). This anti-correlation likely arises from the magnetic moments of $Mn_{Bi}$ ions, which couple antiferromagnetically to the primary Mn layers[59,60]. A high density of magnetic $Mn_{Bi}$ ions suppresses the total magnetization experienced by the electrons, thereby degrading the QAH quantization. When an external magnetic field aligns the magnetic moments of $Mn_{Bi}$ ions with the majority moments, The total magnetization is effectively enhanced. This mechanism explains the pronounced, linearly increasing background in the field-dependent $\Delta E$ below 8 T in earlier samples with relatively low quality[7].

As a strong external magnetic field polarizes the magnetization of all five SLs, the magnetic states in both Type A and Type S devices converge to the FM state (Fig. 1e and 1f). The strong magnetic field, meanwhile, drives the two-dimensional electrons of the FM state into the extreme quantum limit, presenting a unique opportunity to probe a topological system in the presence of quantized LLs. Type S devices are particularly suited to this study due to the high mobility of their surface electrons in the FM state, even though zero-field quantization is not observed in these devices. Indeed, a plethora of quantized topological states emerge as the carrier concentration is modulated under strong magnetic fields up to 45 T. Fig. 2a presents the Hall resistance as a function of gate voltage for a typical 5-SL Type S device (Device S1) under magnetic fields ranging from 15 T (blue) to 42 T (red). Prominent Hall plateaus appear at $h/\nu e^2$, where $\nu = +1, -1, -2, -3$, and $-4$ are integer filling factors in the usual quantum Hall language, accompanied by corresponding minima in $R_{xx}$ (Fig. 2b). Among our observed quantized states, the $\nu = -1$ state is recognized as a QAH state adiabatically connected to the QAH state at zero magnetic field (Fig. 1e). Measurements at elevated temperatures reveal a large, field-dependent excitation gap for the $\nu = -1$ and $\nu = -2$ states, on the order of 100 K—significantly larger than the QAH gap obtained in Type A devices. Finite excitation gaps are also observed at higher filling factors. These findings corroborate the enhanced mobility of Type S devices in the FM state.

These quantized states are further resolved as plateaus in the gate-voltage-dependent Hall conductivity $\sigma_{xy}$ shown in Fig. 2c. The $\sigma_{xy}$ plots additionally reveal an intriguing fractional state (FS) between $\nu = -2$ and $\nu = -3$, which emerges above a threshold magnetic field of 40 T (Fig. 2c, dashed box); closer examination under fields up to 45 T confirms the existence of this state (Fig. 2c upper inset). While the precise nature of this novel fractional state remains unclear, its observation highlights the increasingly significant role of many-body interactions in higher-quality devices.

Fig. 3a displays the longitudinal conductivity $\sigma_{xx}$ as a function of both back-gate voltage $V_g$ and magnetic field $\mu_0 H$, illustrating the intricate topological phases in 5-SL MnBi$_2$Te$_4$. These phases manifest as regions of minimum $\sigma_{xx}$ (with quantized $\sigma_{xy}$, not shown) separated by sharp boundaries. The phase boundaries correspond to LLs in the conventional quantum Hall framework, where electron levels exhibit positive slopes and hole levels exhibit negative slopes, forming the characteristic Landau fan diagram. These LLs are more clearly resolved in the first derivative of $\sigma_{xy}$, $\partial\sigma_{xy}/\partial V_g$, plotted against both $V_g$ and $\mu_0 H$ in Fig. 3b. This derivative plot distinctly resolves the fractional topological phase between the $\nu = -2$ and $\nu = -3$ phases, and uncovers additional topological states at $\nu = -6$ and $-8$. These higher filling factors are determined from nearly quantized $\sigma_{xy}$.

The Landau fan diagram of the 5-SL MnBi$_2$Te$_4$ resembles that of a conventional ambipolar 2D system hosting both electron and hole LLs. However, closer inspection of Fig. 3a reveals major deviations: the lowest levels around the CNP ($L_0$, $L_1$ and $L_2$) do not converge to a single point at zero magnetic field, and one level ($L_2$) intersects multiple other LLs with higher filling factors.

How can we understand this unconventional Landau fan diagram? Important insights arise from analyzing the electronic structure of a thin slab of hypothetical FM MnBi$_2$Te$_4$. Since FM MnBi$_2$Te$_4$ is theorized to be a Weyl semimetal[2,4], a thin slab effectively acts as a 2D quantum well of such a semimetal. Fig. 3c illustrates the calculated $\Gamma$-point energies of sub-bands formed in this 2D quantum well as slab thickness is varied. When the thickness exceeds 2 SLs, an inversion of the surface bands transitions the system from a $C = 0$ trivial insulator into a $C = -1$ QAH insulator. As the thickness increases further, electron and hole bulk bands begin to populate the QAH gap. In a 5-SL slab, three bulk bands ($E_1^e$, $E_1^h$ and $E_2^h$, where the superscript e and h denote electron and hole band, respectively; Fig. 3c) emerge between the pair of surface bands ($E_s^e$ and $E_s^h$), effectively reducing the QAH gap from 88 meV to 39 meV, the energy gap between the $E_1^e$ and $E_1^h$ sub-bands.

A finite magnetic field collapses each of these sub-bands into a set of discrete LLs. Fig. 3e presents the calculated energy spectra of these LLs as they evolve with increasing magnetic field. At 45 T, the lowest three LLs are identified as the first LLs of three distinct sub-bands: $L_0$ originates from the hole sub-band $E_{1,k}^h$, while $L_1$ and $L_2$ derive from the electron bands $E_{1,k}^e$ and $E_{2,k}^e$, respectively (Fig. 3e, right panel). Notably, the $L_2$ level intersects with other LLs at intermediate magnetic fields. These findings provide a compelling explanation for the outstanding features observed in the LL fan diagrams shown in Fig. 3a and 3b. Indeed, density of states (DOS) simulations based on calculated spectra reproduce a fan diagram that closely aligns with the experimental results in Fig. 3b, once level broadening and impurity states are taken into account (Fig. 3d).

With the phase boundaries clarified, we now turn to the topological nature of these phases. Generally, when an external magnetic field is applied, the nontrivial topology of a QAH insulator becomes encoded in its discrete LLs. This QAH topology is fundamentally tied to the parity anomaly of Dirac fermion in $2 + 1$ dimensions[16,15,17]. The LL-resolved topological phase diagram of 5-SL MnBi$_2$Te$_4$ thus provides a unique opportunity for probing this anomaly under a quantizing magnetic field[18,19].

We begin with a phenomenological description of the topological phases. At the CNP, a QAH insulator is characterized by a nontrivial Chern number $C_0$ (for 5-SL MnBi$_2$Te$_4$, $C_0 = -1$ by definition) under zero magnetic field. Our experiments on 5-SL MnBi$_2$Te$_4$ show that the QAH gap remains open under finite magnetic fields (Fig. 1e). This observation implies that the same $C_0$, through adiabatic continuity, continues to describe the QAH phase in the presence of LLs. Since each LL carries a Chern number of $-1$, the system's total Chern number decreases by $-1$ (or increases by $+1$) each time the Fermi level crosses an initially unoccupied (or occupied) LL. Consequently, the topological phases of the system at finite doping are characterized by the topological number $N = C_0 - \Delta N_+ + \Delta N_-$, where $\Delta N_+$ ($\Delta N_-$) represents the number of initially unoccupied (occupied) levels that the Fermi level traversed from the CNP. This topological number, $N$, as labeled on the phase diagram in Fig. 3d, accurately describes the quantized Hall conductivity, $\sigma_{xy} = Ne^2/h$, observed in these phases.

A deeper understanding of the topological phases invokes the parity anomaly of Dirac fermion in $2 + 1$ dimensions. This anomaly reflects the intrinsic conflict between time-reversal symmetry and gauge invariance when regularizing the Dirac fermion theory in $2 + 1$ dimensions[17]. In QAH systems such as 5-SL MnBi$_2$Te$_4$, the appropriate regularization scheme breaks time-reversal symmetry while preserving gauge symmetry[17]. Under this scheme, the parity anomaly manifests as a topological quantity known as the spectral asymmetry $\eta$, which quantifies the asymmetry of the entire spectrum[18,19]. In the presence of LLs with uniform degeneracy, $\eta$ corresponds to the imbalance between the total number of unoccupied and occupied LLs. This understanding allows the topological number $N$ to be reformulated as a generalized topological index:

$$N = \frac{N_+ - N_-}{2},$$

where $N_+$ ($N_-$) is the total number of unoccupied (occupied) LLs. Comparing this expression to the earlier definition of $N$ yields $N_- = N_+ - 2C_0$ at the QAH gap, a universal relation that encapsulates the parity anomaly for all QAH systems under external magnetic fields.

In 5-SL MnBi$_2$Te$_4$, this relation implies that, counted from the QAH gap, the occupied LLs outnumbers the unoccupied LLs by 2. (The fact that these LLs originate from different sub-bands in 5-SL MnBi$_2$Te$_4$ does not affect the analysis.) Compared with a topologically trivial insulator, where $N_- = N_+$, the QAH topology (i.e., the parity anomaly) of 5-SL MnBi$_2$Te$_4$ can be viewed as an electron LL inverted to the hole side. Indeed, our calculations identify an anomalous LL—the first LL of the surface band $E_{s,k}^h$, as shown in Fig. 3e—that is responsible for the $\nu = -1$ QAH effect observed in 5-SL MnBi$_2$Te$_4$. This anomalous LL, although originating from a hole band, disperses as an electron level all the way to the $\mu_0 H \to \infty$ limit (Supplementary Fig. 6). So, unlike other hole LLs, this singular level curls upwards from below and crosses the Fermi level at the sample boundary. The resulting chiral edge state gives rise to the quantized QAH conductance that we observe experimentally at $\nu = -1$. This and the other quantized topological states in MnBi$_2$Te$_4$, therefore, go far beyond a simple superposition of quantum Hall and QAH effects[7], and are fundamentally distinct from the 2D Dirac cone models proposed in ref.[61].

Our calculations further predict exotic edge transport in the $\nu = 0$ phase. A schematic of the anticipated LL structure in this phase is presented in Fig. 4b. The upward-dispersing anomalous LL intersects a downward-dispersing normal hole level (the first LL of the $E_{1,k}^h$ sub-band) at the edges. The level crossings create two counter-propagating edge channels on both sides of the specimen, resembling the scenario in a quantum spin Hall insulator[62–64]. The backscattering and localization likely opens a transport gap at their crossing points[65,66], dividing the transport into three regimes, referred to as $0^-$, $0'$, and $0^+$, as the Fermi level shift through the $\nu = 0$ phase. The edge channels in these regimes are illustrated in Fig. 4c. Remarkably, three distinct regions, attributable to the $0^-$, $0'$, and $0^+$ regimes, are evident in both $\sigma_{xx}$ and $\sigma_{xy}$ as the gate voltage sweeps through the $\nu = 0$ phase. These regimes are even more distinctly resolved in the phase diagrams in Fig. 3a and 3b, where sharp boundaries separate the three regimes. These findings provide additional support for our understanding of the topologically nontrivial electronic structure of 5-SL MnBi$_2$Te$_4$.

To conclude, we have fabricated devices of intrinsic magnetic topological insulator MnBi$_2$Te$_4$ with much improved quality. The advancement enables better QAH quantization at zero magnetic field. More importantly, the high-quality samples produce a rich set of quantized topological states under high magnetic fields, uncovering fundamental links between parity anomaly, anomalous LL structure, and QAH topology. With this longstanding challenge in MnBi$_2$Te$_4$ research now overcome, we anticipate that this material will serve as a major playground for exploring novel topological quantum physics.

**Acknowledgements**

We thank Philip Kim, Xiaofeng Jin, Yizheng Wu, Dung-Hai Lee, Taige Wang, Zhiqiang Gao, Raquel Queiroz and Yijun Yu for helpful discussions. Part of the sample fabrication was performed at Nano-fabrication Laboratory at Fudan University. A portion of this work was performed at High Magnetic Field Laboratory, HFIPS, Chinese Academy of Sciences, Hefei, China. Z.G., J.G., Z.H., D.Y., M.L., W.R., and Y.Z. acknowledge support from National Key R&D Program of China (grant no. 2022YFA1403301), National Science Foundation of China (grant no. 1235000130), Innovation Program for Quantum Science and Technology (grant no. 2024ZD0300104), and Shanghai Municipal Science and Technology Commission (grant nos. 23JC1400600 and 2019SHZDZX01). W.R. acknowledges support from National Science Foundation of China (grant no. 12350401) and Shanghai Science and Technology Development Funds (grant no. 22QA1400600). W.R. and H. J. acknowledges support from National Science Foundation of China (grant no. 12350401). J.G. acknowledges support from the National Natural Science Foundation of China (grant no. 12204115) and China Postdoctoral Science Foundation (grant no. 2022M720812). D.Y. acknowledges support from the National Natural Science Foundation of China (grant no. 12004075) and China Postdoctoral Science Foundation (grant no. BX20190084 and 2019M661330). G.H. and C.Z. acknowledge support from National Science Foundation of China (grant no. 62171136). Z.L. and J.W. acknowledge support from National Key R&D Program of China (grant nos. 2019YFA0308404), National Science Foundation of China (grant nos. 12350404, 12174066), Shanghai Municipal Science and Technology Commission (grant no. 23JC1400600, 2019SHZDZX01). C.X. and G.K. acknowledge support from Systematic Fundamental Research Program Leveraging Major Scientific and Technological Infrastructure, Chinese Academy of Sciences under contract no. JZHKYPT-2021-08. Y.S. acknowledges support from National Key R&D Program of China (grant no. 2021YFA1600201), National Science Foundation of China (grant no. 12274412). This work has been supported by the New Cornerstone Science Foundation through the New Cornerstone Investigator Program and the XPLORER PRIZE.

**Contributions**
Y.Z., W.R., J.W., and X.H.C. supervised the project. J.G. and Z.G. grew the bulk crystals, with assistance from M.S. and Z.X. Z.G. and J.G. fabricated few-layer devices and conducted transport measurements. Z.H. carried out the STM measurements. Z.G., J.G., and D.Y. performed high-field transport measurements at HFIPS, with support from C.X. and G.K. S.W. (Shuang Wu), X.C., and S.W. (Shiwei Wu) performed RMCD measurements. G.H. and C.Z. performed TEM measurements. Z.G., J.G., D.Y., Z.L., J.W., W.R., and Y.Z. analyzed the data. Z.L., H.J., X.C.X., and J.W. performed theoretical calculations. Z.G., W.R., and Y.Z. wrote the manuscript, with input from all authors.

**Figures**

**Figure 1 | QAH effect and magnetic transitions in 5-SL MnBi₂Te₄ devices. a**, Schematic of the device structure in Type A configuration, overlaid with an optical image of Device A1. The fully transparent substrate comprises $Al_2O_3$/ITO layers supported on a 500-μm-thick quartz wafer, with the ITO layer serving as a back gate. Contacts for standard four-terminal $R_{xx}$ measurements are marked. Scale bar, 10 μm. **b**, Temperature dependence of $R_{xx}$ for four representative Type A 5-SL devices. The Néel transition is evident as kinks in $R_{xx}$ at $\sim 23.5$ K. **c**, $R_{yx}$ (upper panel) and $R_{xx}$ (lower panel) of Device A1 as functions of magnetic field at $T = 1.5$ K. The hallmark QAH effect is characterized by quantized $R_{yx}$ and vanishing $R_{xx}$ at $\mu_0 H = 0$ T. **d**, $R_{yx}$ (upper panel) and $R_{xx}$ (lower panel) of Device A1 as functions of magnetic field at various temperatures. All data in (c) and (d) are acquired at the CNP ($V_{CNP} = -20$ V), with $R_{yx}$ antisymmetrized and $R_{xx}$ symmetrized to eliminate the mixing of the two components. $R_{yx}$ is additionally rescaled to correct for a 1.01% systematic error in our resistance measurements. **e**, QAH excitation gap versus magnetic field for Devices A1 and A2, derived by fitting the Arrhenius plots of temperature-dependent $R_{xx}$. Shaded regions represent the error bound of the extracted energy gaps. As the magnetic field increases from zero, the sample transitions through four magnetic states: AFM, M3′, CAFM, and FM. Insets depict the schematic magnetic configurations of these states. **f**, First derivative of $R_{xx}$, $\partial R_{xx}/\partial(\mu_0 H)$, as a function of magnetic field and temperature, measured in a 5-SL device in the Type S configuration (Device S1). Below the Néel temperature, the same magnetic states as in (e) appear as distinct regions separated by magnetic transitions. The sample enters the paramagnetic (PM) state above the Néel temperature. Magnetic transitions are marked by colored ticks. The same set of ticks also indicate transitions in (c), (d), (e), and (f).

**Figure 2 | Quantized topological states in 5-SL MnBi₂Te₄ under high magnetic fields. a-b**, Gate-dependent $R_{yx}$ (a) and $R_{xx}$ (b) measured in Device S1 under magnetic fields from 15 T to 42 T at $T = 0.3$ K. Quantized Hall plateaus in $R_{yx}$ are observed at $h/\nu e^2$, where $\nu = +1, -1, -2, -3, -4$ (dashed lines in a). These plateaus are accompanied by vanishing $R_{xx}$ in (b), as highlighted by vertical lines and the associated filling factors $\nu$. **c**, Gate-dependent Hall conductivity $\sigma_{xy}$, derived from $R_{yx}$ and $R_{xx}$ in (a) and (b), under various magnetic fields. Quantized plateaus are clearly resolved at filling factors $\nu = +1, -1, -2, -3, -4$ (dashed lines), along with the emergence of a developing plateau at a fractional filling between $\nu = -2$ and $\nu = -3$ (dashed box). Upper inset: Magnified view of the fractional state as it evolves under magnetic fields up to 45 T. Lower inset: Schematic of the device structure in the Type S configuration, overlaid with an optical image of Device S1.

**Figure 3 | Topological phase diagram of 5-SL MnBi₂Te₄. a**, $\sigma_{xx}$ as a function of gate voltage and magnetic field, measured in Device S1 at $T = 0.3$ K. The topological phases are labeled by their corresponding filling factors $\nu = +1, 0, -1, -2, -3, -4$. $L_0$, $L_1$, and $L_2$ denote the lowest three LLs near the CNP. The $\nu = 0$ phase can be divided into three distinct regimes, separated by sharp boundaries, and labeled as $0^-$, $0'$, and $0^+$. **b**, Derivative of the Hall conductivity, $\partial \sigma_{xy}/\partial V_g$, as a function of gate voltage and magnetic field, measured in the same device in (a) at $T = 0.3$ K. In addition to the topological phases observed in (a), additional phases are resolved at $\nu = -6, -8$, along with a fractional phase between $\nu = -2$ and $\nu = -3$ phases. Dashed lines in (a) and (b) serve as guides to the eye. **c**, Calculated Γ-point energies of

sub-bands in a hypothetical 2D quantum well of ferromagnetic MnBi$_2$Te$_4$ at zero magnetic field, as a function of the quantum well thickness, $d$. The inversion of the two surface bands occurs as $d$ exceeds 2-SL, transitioning the system from a trivial insulator ($C = 0$) to a QAH insulator ($C = -1$). As $d$ further increases, bulk bands begin to populate the gap between the surface bands. The $\Gamma$-point energies of 5-SL MnBi$_2$Te$_4$ are labeled as $E_n^{e/h}$, where $e/h$ indicates electron or hole sub-bands, and $n = s, 1, 2 \ldots$ labels the surface and bulk bands. **d**, Simulated DOS based on calculated LL spectra of 5-SL MnBi$_2$Te$_4$, incorporating LL broadening and impurity states. Details of the simulation are presented in Fig. S5. **e**, Calculated field-dependent LL spectra (left panel), schematic band structure (middle panel) and high-field LL DOS (right panel) for 5-SL MnBi$_2$Te$_4$. Sub-bands are labeled as $E_{n,k}^{e/h}$, where $k$ is the wavevector. The first LL of the $E_{s,k}^{h}$ sub-band is identified as an anomalous LL—a hole level that disperses like an electron level—responsible for the QAH topology and parity anomaly in 5-SL MnBi$_2$Te$_4$ under high magnetic fields (see text).

**Fig. 4 | Helical edge transport in the $\nu = 0$ phase. a**, Gate-dependent $\sigma_{xx}$ (left panel) and $\sigma_{xy}$ (right panel) in the $\nu = 0$ phase under high magnetic fields ranging from 32 T to 42 T. Three distinct regimes, labeled as $0^-$, $0'$, and $0^+$, are identified within this phase. **b**, Schematic LL structure of a finite 5-SL MnBi$_2$Te$_4$ in the $\nu = 0$ phase. The anomalous LL (the first LL of the $E_{s,k}^{h}$ sub-band, shown in red) intersects with a hole LL (the first LL of the $E_{1,k}^{h}$ sub-band, shown in blue) at the two edges of the sample, forming two counter-propagating helical edge channels on each edge. Backscattering between the two channels likely induces a hybridization gap at the level crossing. Edge transport is thus divided into three scenarios, corresponding to $0^-$, $0'$, and $0^+$, as the Fermi level shift through the $\nu = 0$ gap. **c**, Real-space illustration of the edge transport anticipated in the $0^-$, $0'$, and $0^+$ regimes. In the $0^-$ and $0^+$ regimes, red and blue lines represent the two counter-propagating chiral edge channels arising from LL crossing in (b). These edge channels are absent in the $0'$ regime.

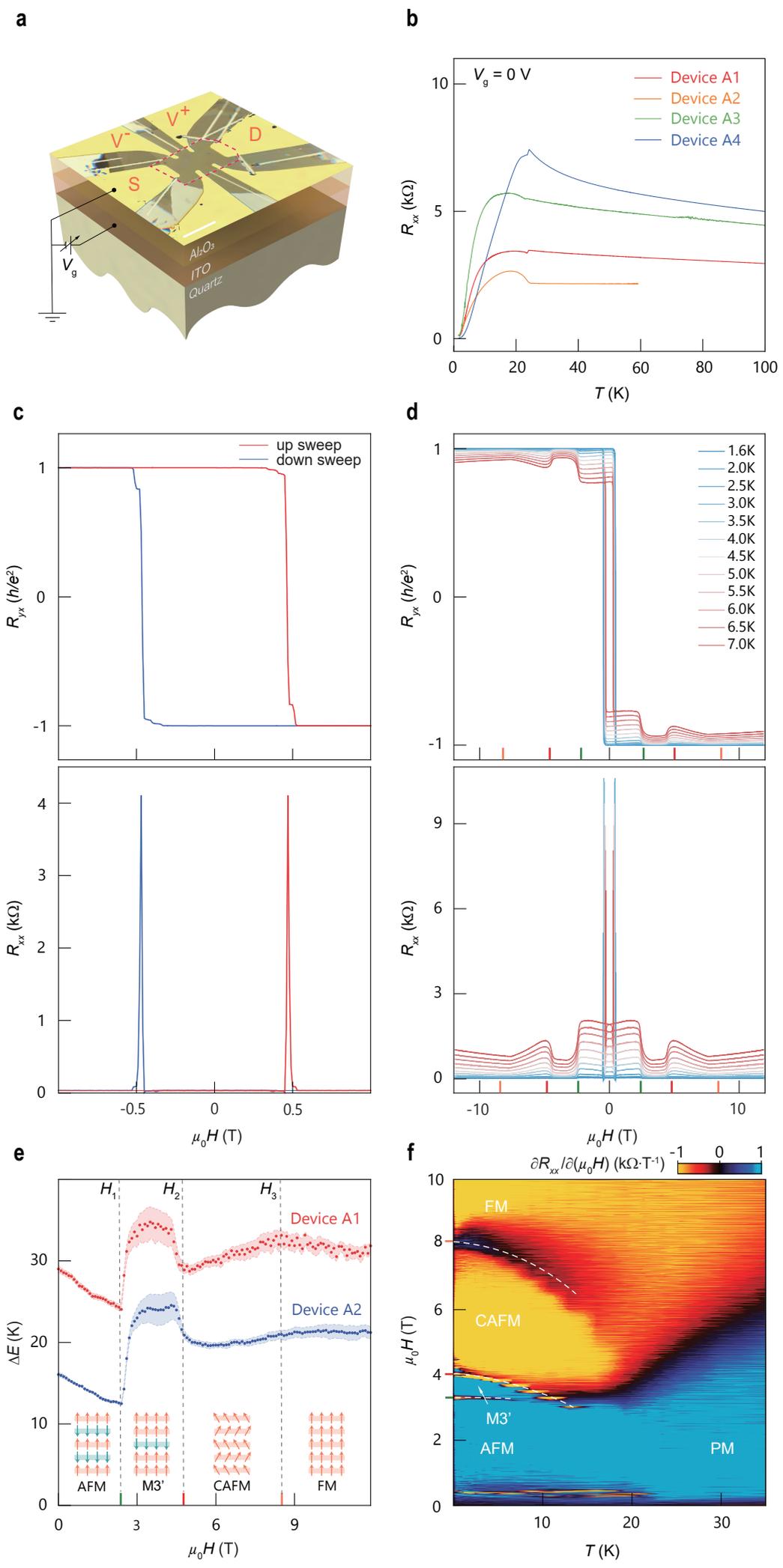

Fig. 1, Z.X. Guo *et al*

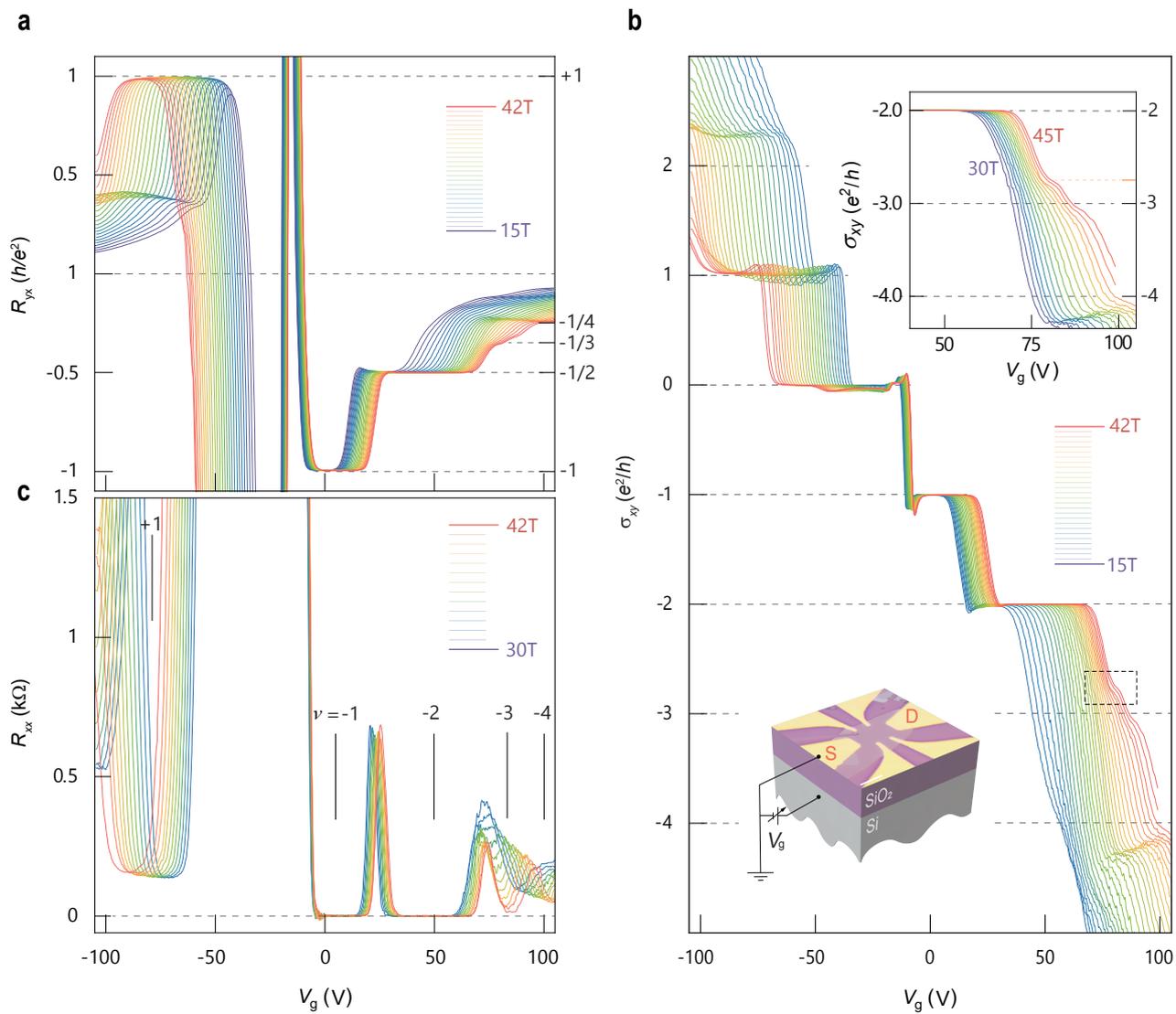

Fig. 2, Z.X. Guo *et al*

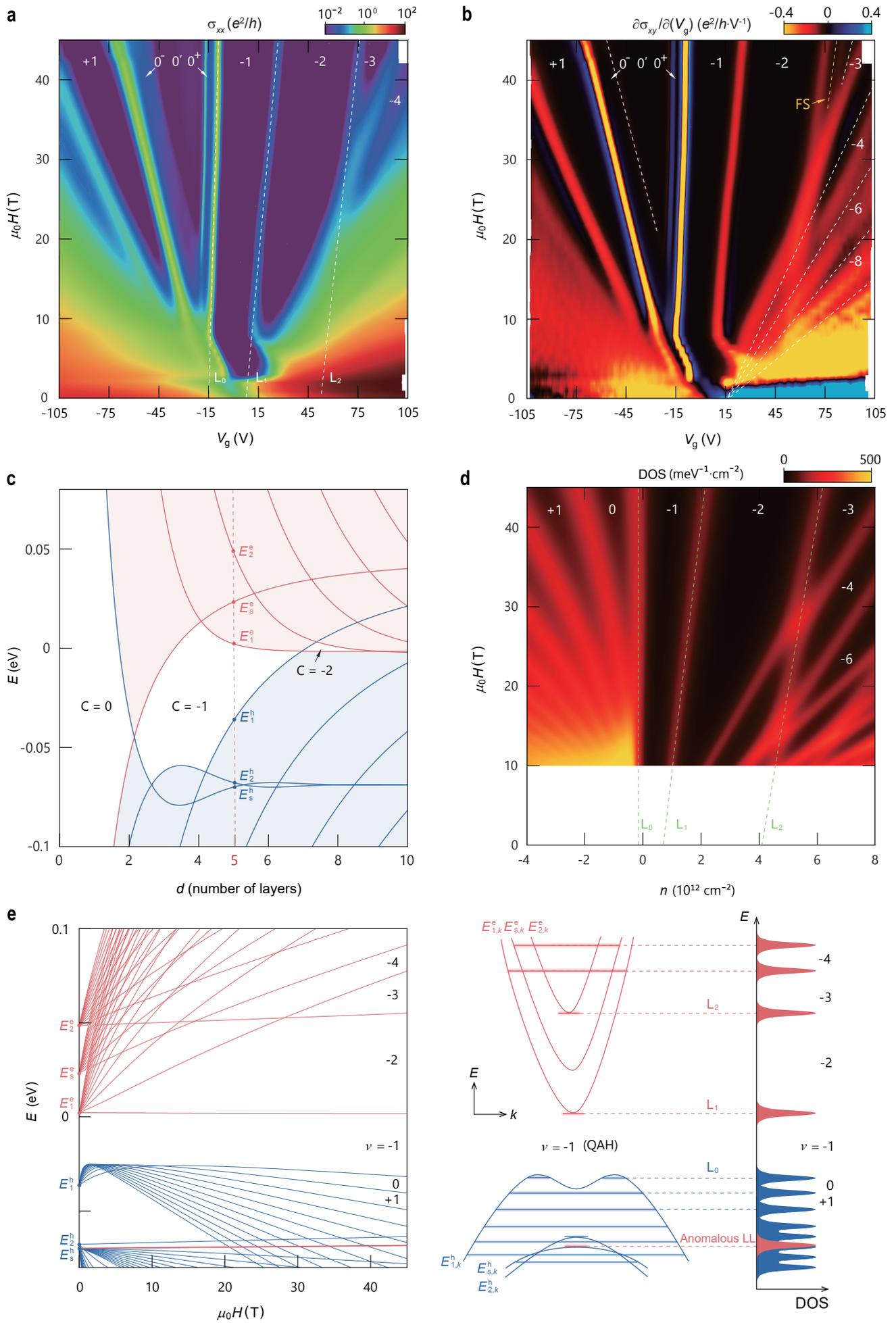

Fig. 3, Z.X. Guo *et al*

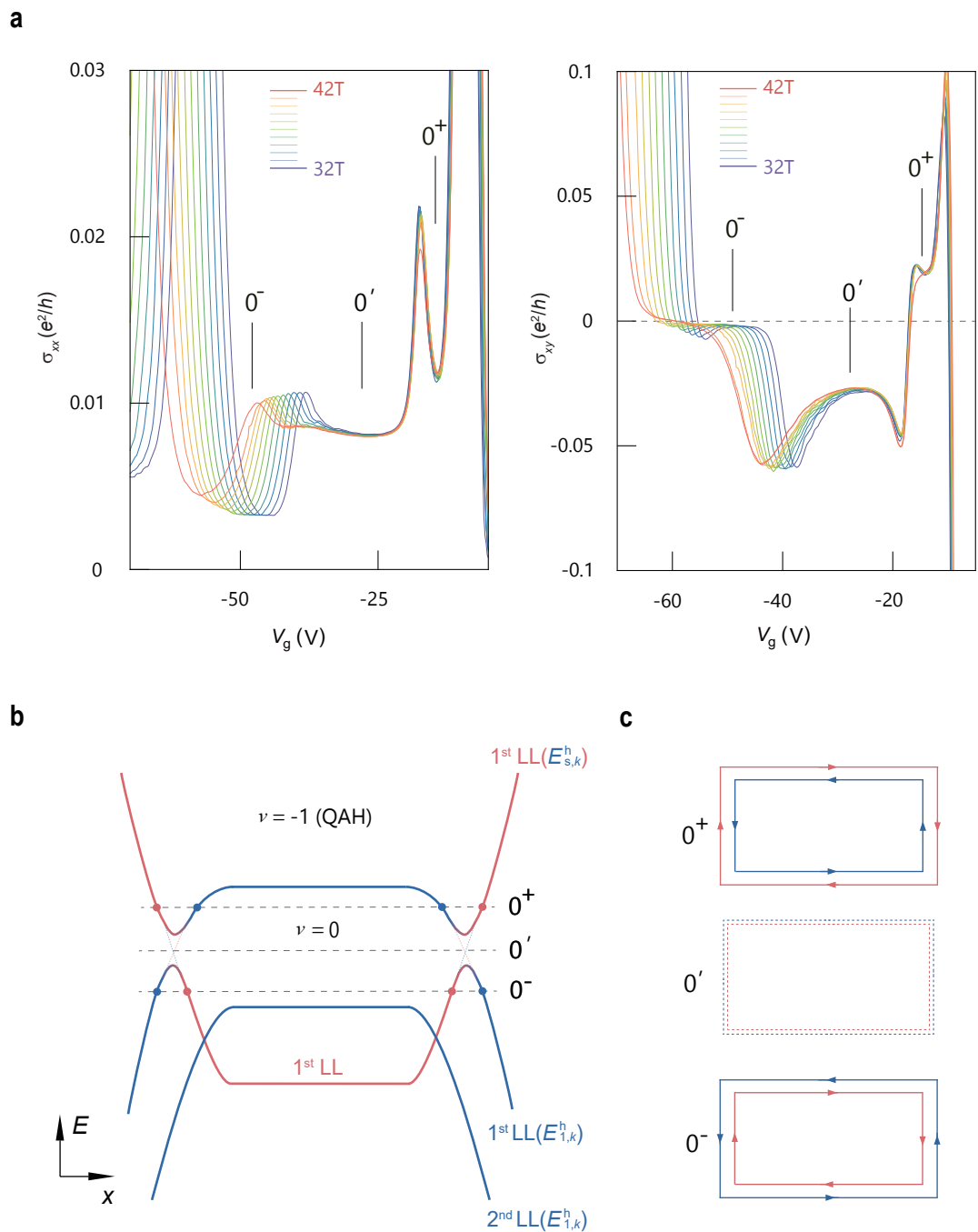